# How does certainty enter into the mind?

Ching-an Hsiao

*h_siao@hotmail.com*

**Any problem is concerned with the mind, but what do minds make a decision on? Here we show that there are three conditions for the mind to make a certain answer. We found that some difficulties in physics and mathematics are in fact introduced by infinity, which can not be rightly expressed by minds. Based on this point, we suggest a general observation system, where we use region (a type of infinity) to substitute for infinitesimal (another type of infinity) and thus get a consistent image with the mind. Furthermore, we declare that without world pictures we can never have ideas to any expressive events, which is the primary condition for a wave function like mind to collapse to a series of numbers. A following observation by expanding algorithm brings the final collapse: classifying the numbers and coming up with a certain yes or no answer.**

There is no bigger question than the mind one in nature. All scientific, economic and cultural actions are exerted by the minds and in the end for the minds. The meaning of nature lays in its being sensed. Anything can not be sensed or observed is out of science scope. One interesting problem about mind comes from experiment performed by Libet: time-delays of consciousness [1]. The experiment includes two parts. The first part shows that it took about half a second before a stimulus applied to the skin to be consciously aware of, but no subjective delay feeling can be sensed. And the second part says that the order of stimuli on the skin and corresponding point of the somatosensory cortex are false perceived. A late stimulus on the skin was sensed as occurred before a former one of the cortex. The astonishing delay caused Libet to believe that there is no free will. We will not discuss whether there is free will here, we focus on how minds sense. But to the experiment, we have two certain points. One is that time order is not so absolute to mind (it does match relativity theory), the other is time threshold is necessary to cause a sensation.

It is undoubted that minds can perceive events within a certain time and space. But can minds locate infinitesimal? Or, is there any sensible infinitesimal? It seems no. On one side, we see thresholds everywhere, especially to the mind. One early work in this field by Weber [2] leaded to the well-known Weber's Law, which tells us that noticeable difference is limited. On the other side, the paradoxes of motion of Zeno [3] remind us that there is no sensible (measurable) infinitesimal. The Achilles and the Dichotomy paradoxes show there is no sensible infinitesimal space, and the arrow paradox shows no sensible infinitesimal time. Because to a certain number in continuum, one can't find its certain nearest neighbour, continuum time-space is in fact not successive and then can not be observed. In this way, we have to set up our observing system not only on relativity but also on quantum, i.e. any reference system should take a limited time-space unit. Motion can thus be definitely expressed as space difference with time difference. In such a system time-space becomes successive and Achilles can also overtake the tortoise easily. By introducing this system, we find when a "photon" is incident upon a double slit it might never plan to pass by only one. The case is probable that by observing, it collapses (with original state in continuum) to one photon (Copenhagen interpretation) or we enter into different worlds (MWI), etc. In latter case, Zeno's arrow seems "rest" indeed as he expected. Uncertainty is the innate nature of the system.

Now we got the first condition for certainty entering into mind: reference systems with respective limited units. Recent research [4, 8] indicates other conditions. One is the picture of world [5]. Precious research [4] found that without specified patterns, we can never tell the "abnormal" from the "normal". Generalized conclusion is that no pattern can be identified absolutely. Following is a mathematical example.

To Cantor's theory of transfinite number, Hilbert declared, "No one shall expel us from the paradise that Cantor has created" [6]. Yes, it is quite attractive, and yet controversial. In this theory, the diagonal method fascinates many persons, and calls criticisms simultaneously. Wittgenstein reviewed with strong confidence that "There is always one of the series for which it is not determined whether or not it is different from the diagonal series" and "it may be said: they run after one another to infinity, but the original series is always ahead" [7]. Since כ Algorithm [4] tells us that certainty is indeed dependent on world picture, let us try to find foundations of both. Cantor believed that we can construct an inverted diagonal series but Wittgenstein did not. It is not too enough to praise the elegance of the diagonal method, but we should not neglect what Wittgenstein said. To real numbers in (0, 1), we have for sure such a case, where exists one-to-one correspondence matched with natural numbers. The numbers are ordered by length of digits and size of numbers as following.

{0.1, 0.2,…, 0.9, 0.01, 0.02,..., 0.09, 0.11, 0.12,…, 0.19,…, 0.99, 0.001, 0.002,…… }

The diagonal sequence is 0.1000000…. (same as the first number). Perfectly as Wittgenstein imagined, we don't know the exact position of any inverted diagonal sequence (such as 0.01111…..). If one thinks that 0.1 should be expressed by 0.09999…., result is the same. Different world pictures tell us different things, it does be what כ Algorithm shows us. Is $c$ really greater than $\aleph_0$? We were ever in illusion.

The third condition for mind certainty is the mechanism of collapse. Any signals into the mind can be reasonably expressed by numbers in the end, but how to split (classify) the numbers is not yet certain. We now have expanding algorithm for it [4]. Expanding algorithm is based completely on inside structure of data – consistence or inconsistence. A sensitive index called Integrated Inconsistent Rate (IIR) was introduced in this algorithm, which is a subjective parameter and can be deduced by Weber's law. Weber's law states that the ratio of the increment threshold ($\Delta I$) to the background intensity ($I$) is a constant ($K$), i.e. $\frac{\Delta I}{I} = K$. All distinguishable quantity is related to this formulation. Basically, let us give three values deduced from the formulation: $0, I, I + \Delta I$. When we can not tell $I$ from $I + \Delta I$, number 0 is a distinguishable quantity. Or, we can make a transformation: $0, \Delta I, I + \Delta I$. When $\Delta I$ can not be sensed, $I + \Delta I$ is different (sensible) from them (0 and $\Delta I$). We define three quantities for value $I + \Delta I$. Expansion ratio expresses an expanding degree of interval of two neighbours to average interval of the whole, inhibitory rate is a factor to modify expanding effect by comparing current interval and former interval.

Expansion ratio: $Er = \frac{I + \Delta I - \Delta I}{(I + \Delta I)/2} = \frac{2}{1+K}$

Inhibitory rate: $Ihr = \frac{I}{I - \Delta I} = \frac{1}{1-K}$

Integrated Inconsistent Rate: $IIR = \frac{Er}{Ihr} = \frac{2 \times (1-K)}{1+K}$

We have reasonable $K$ in (0, 1), $K = 0$ means sensing infinitesimal, $K = 1$ means no sense and $K > 1$ means negative sense. We have corresponding typical *IIR* in Table 1. They are also the information parameters being taken with a reference system. Effect of expanding algorithm can refer to [4, 8]. Another interesting example is the explanation to Cornsweet Illusion (Fig. 1), which perfectly demonstrates the "collapse" or consistent interpretation of minds to relative sensible quantities. In the image, major brightness is 173 (0 is black and 255 is white). Related values of two sides are listed by pixel as a set C{173, 172, 171, 170, 169, 168, 167, 178, 177, 176, 175, 174, 173}. The narrow band from 172 to 167 and 178 to 174 in the middle brings different bright feeling to same most 173s on both sides. By running Expanding Algorithm to set C, we can divide them into two parts easily. One part is from 173 to 167, the other is from 178 to 173.

One function of mind is possible to construct relationship between events by time and space [8]. Return to Libet's time-delays of consciousness, can we affirm that there is no mind before the end of the half a second? How is it if we say minds get information continuously from the beginning of the half a second to the end of the half a second, collapse at the end of the half a second and respond a complete and consistent picture? To skin stimulus, it is necessary for minds to build a time correspondence with the beginning, while to the cortex, minds choose the end experientially or transcendentally. After all, according to relativity, there is not really such a thing as the 'now' at all. So except for sensible related events, time and space have no absolute order to mind. What if mind constructs changeable relations according to experience?

Shakespeare was right about the world being a stage and we being actors, which might be the meaning of the minds. We experience the world with self information system (limited units, world picture and sensible index), and then change the world and even ourselves. We always have certain answers, about it we never doubt. After all, how can one be wrong to what he actually percepts? Can we know more than what perception tells us? Like the French lady mentioned in Benjamin Franklin's speech "on the faults of the constitution", how could our mind perceive things out of our minds?

However, mind problem still puzzles our minds (like free will problem). It is interesting that we have to say: if one doesn't puzzle on mind problem, he has not understood the mind.

**Table 1 Typical *IIR* in three values system**

| K | IIR |
|---|---|
| *0* | 2 |
| 0.01 | 1.96 |
| 0.05 | 1.81 |
| 0.1 | 1.64 |
| 1 | 0 |

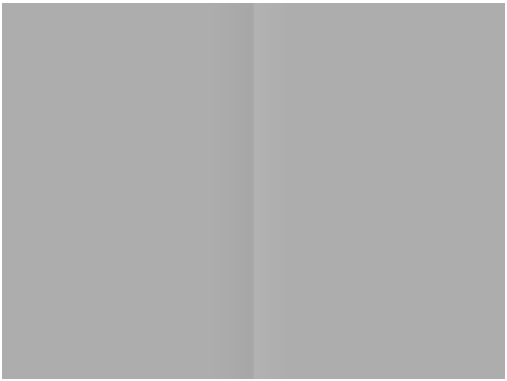

**Figure 1: Cornsweet illusion.** Left part of the picture seems to be darker than the right one. In fact, there have the same brightness as can be seen by blanking out the middle region containing the edge.